\documentclass[useAMS,usenatbib]{mn2e}
\usepackage{epsfig,lscape} 
\usepackage{natbib}
\usepackage{amsmath}
\usepackage{color}

\newcommand{\ha}{\hbox{H$\alpha$}}
\newcommand{\hb}{\hbox{H$\beta$}}

\newcommand{\oiii}{\hbox{[O\,{\sc iii}]}}

\newcommand{\nii}{\hbox{[N\,{\sc ii}]}}

\title[Modelling the mass-metallicity relation of star-forming galaxies]{Modelling the mass-metallicity relation of star-forming galaxies from $z\sim 3.5 $ to $z\sim 0$}

\author[J. Lian et al.]
{Jianhui Lian\thanks{jianhui.lian@port.ac.uk},
	Daniel Thomas, Claudia Maraston\\
	Institute of Cosmology and Gravitation, University of Portsmouth, Burnaby Road, Portsmouth, PO1 3FX, UK
}

\begin{document}
	
\maketitle

\begin{abstract}
We study the origin and cosmic evolution of the mass-metallicity relation (MZR) in star-forming galaxies based on a full, numerical chemical evolution model. The model was designed to match the local MZRs for both gas and stars simultaneously. This is achieved by invoking a time-dependent metal enrichment process which assumes either a time-dependent metal outflow with larger metal loading factors in galactic winds at early times, or a time-dependent Initial Mass Function (IMF) with steeper slopes at early times. We compare the predictions from this model with data sets covering redshifts $0\leq z\leq 3.5$.
The data suggests a two-phase evolution with a transition point around $z\sim 1.5$. Before that epoch the MZR$_{\rm gas}$ has been evolving parallel with no evolution in the slope. After $z\sim 1.5$ the MZR$_{\rm gas}$ started flattening until today.
We show that the predictions of both the variable metal outflow and the variable IMF model match these observations very well. {Our model also reproduces the evolution of the main sequence, hence the correlation between galaxy mass and star formation rate.} We also compare the predicted redshift evolution of the MZR$_{\rm star}$ with data from the literature. As the latter mostly contains data of massive, quenched early-type galaxies, stellar metallicities at high redshifts tend to be higher in the data than predicted by our model. Data of stellar metallicities of lower-mass ($<10^{11}\;M_\odot$), star-forming galaxies at high redshift is required to test our model.  
\end{abstract}


\section{Introduction}
Metallicity, defined as the fractional abundance of elements heavier than Helium, is a key parameter in galaxy evolution models, playing an important role in a number of fundamental astrophysical processes such as {star formation, gas cooling,} and dust formation. A tight correlation between the metallicity of {ionized} gas {and stellar mass of} star forming galaxies 
has been {found} for decades  \citep{lequeux1979}. This mass-gas metallicity relation (hereafter MZR$_{\rm gas}$) was later confirmed to be valid over three orders of magnitude in galaxy stellar mass through the analysis of large galaxy samples from the Sloan Digital Sky Survey (SDSS, York et al. 2000) using various gas metallicity calibrations \citep{tremonti2004,kewley2008,andrews2013}. More recently, the MZR$_{\rm gas}$ was detected down to the very low mass end ($M_*<10^9{\rm M_{\odot}}$, \citealt{henry2013a,henry2013b,lian2016}).
Furthermore, based on deep spectroscopy of distant galaxies, several studies found that the relation was already well established at high redshift since at least $z\sim3.5$ \citep{savaglio2005,erb2006,maiolino2008,yabe2013,ly2016,kashino2017}. Compared to local star forming galaxies, high redshift objects at the same stellar mass are generally found to be more metal poor in the ionized gas by $\sim0.3$ dex at $z\sim2$ \citep{erb2006}.

However, the exact evolution of the MZR$_{\rm gas}$ with time and the physical drivers of such evolution is a matter of debate. 
Using the `Te' method to derive gas metallicity, \citet{ly2016} find no clear evolution {in} the slope of the MZR$_{\rm gas}$ since $z\sim1$. \citet{kashino2017}, instead, in a study based on large galaxy samples from the FMOS-COSMOS survey find the MZR$_{\rm gas}$ to be significantly steeper at $z\sim1.6$ than at $z=0$. 
Interestingly, a flattening of the MZR$_{\rm gas}$ relation with cosmic time is consistent with prediction from cosmological hydrodynamical simulations \citep{guo2016,taylor2016}. 

Further constraints on the metal enrichment history of galaxies and hence the evolution of the MZR$_{\rm gas}$ can be obtained from the local mass-{\em stellar} metallicity relation (hereafter MZR$_{\rm star}$), 
since the stellar metallicity carries information on the past metal enrichment. In \citet{lian2018} (hereafter Paper I) we simultaneously analyse the MZR$_{\rm gas}$ and MZR$_{\rm star}$ of local star forming galaxies based on SDSS data and a full, numerical chemical evolution model. Because of the relatively low stellar metallicity of the lower-mass galaxies, we find that these relations can be reproduced simultaneously only if a time-dependent metal enrichment process is invoked where the metal-enrichment in low-mass galaxies is suppressed at early times. We {showed} that 
such a process {could be either metal outflow with higher outflow fractions (i.e. larger metal loading factors) at early times or an IMF with steeper IMF slopes at early times.} 
{The former regulates the {\em metal retention} in galaxies so that more metals are lost at early times, while the latter regulates the {\em metal production} so that less metals are produced at early times.} 

In both scenarios, the cosmic evolution of the MZR$_{\rm gas}$ is mostly driven by the efficiency of star formation in galaxies but further regulated by the cosmic evolution of either the metal outflow loading factor or the IMF slope. The aim of the present paper is to test these predictions for the redshift evolution of the MZR$_{\rm gas}$ using  observational results at various redshifts up to $z\sim3.5$.  

Throughout this paper, we adopt a standard cosmology with parameters with $H_0=71 {\rm km s^{-1} Mpc^{-1}},\ \Omega_{\Lambda}=0.73$ and $\Omega_{\rm m}=0.27$ \citep{spergel2003}.

\section{Observations}
Following Paper I, we select a local sample of star forming galaxies from the SDSS Data Release 12 \footnote{http://www.sdss.org/dr12/} (DR12; \citealt{alam2015}) using the following selection criteria: 
(1) high specific star formation rates, ${\rm log(sSFR)}>-0.6*{\rm log}(M_*/{\rm M_{\odot}})-4.9$;
(2) a stellar mass cut at $M_*>10^9 {\rm M}_{\odot}$; 
(3) a redshift range of $0.02\leq z\leq 0.05$ to ensure mass completeness {up to $10^{11} {\rm M_{\odot}}$};
(4) a high signal-noise-ratio (SNR) above 5 in the strong emission lines (including \hb, \oiii$\lambda\lambda4959,5007$, \ha, \nii$\lambda6584$); 
(5) a classification outside the composite or AGN region through the demarcation of \citet{kewley2001} in the BPT diagram \citep{baldwin1981}. The final sample contains 4633 galaxies. Emission line fluxes are corrected for galactic internal extinction using the Balmer decrement method and the Milky Way extinction law \citep{ccm89}. 
We obtain the gas metallicities of these galaxies using the empirical linear N2 method from \citet{pettini2004}, which is widely-used for determining gas metallicities, especially for high-redshift galaxies. 

\subsection{The mass-gas metallicity relation}
\begin{figure}
	\centering
	\includegraphics[width=\columnwidth,viewport=10 10 580 500,clip]{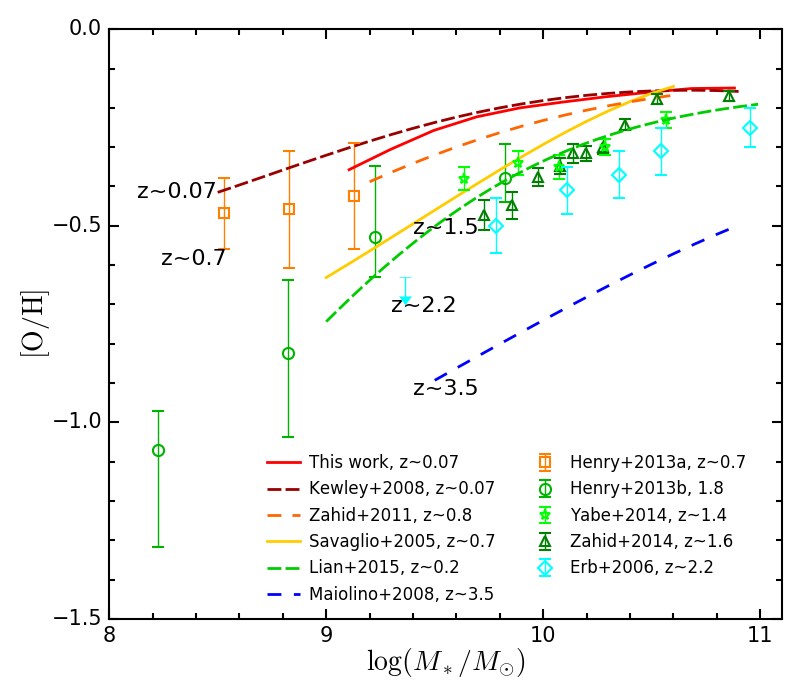}
	\caption{Relation between galaxy stellar mass and gas metallicity at various redshifts. The red solid line is the mass-gas metallicity relation MZR$_{\rm gas}$ at $z\sim0.03$ as derived in this work. The error bars indicate the 1$\sigma$ uncertainty of the metallicity measurements. Gas metallicities are re-calibrated to the empirical N2 method \citep{pettini2004} and stellar masses are corrected to a mass based on a Kroupa IMF.  
}
	\label{figure1}
\end{figure}
Figure 1 shows the local MZR$_{\rm gas}$ we obtain (red line) compared to the same relation at different redshifts as obtained from various works in the literature (other coloured lines as labelled on the plot).
In order to enable a proper comparison, stellar masses are corrected to a stellar mass based on the Kroupa IMF and gas metallicities are re-calibrated to the empirical N2 method using the conversion by \citet{maiolino2008}. 
{We adopt a factor of 0.85 to convert the stellar mass based on Salpeter IMF to that based on Kroupa IMF \citep{maiolino2008,bolzonella2010,pforr2012} and a factor of 1.06 from Chabrier IMF to Kroupa IMF \citep{mannucci2010}.} 
The typical 1$\sigma$ uncertainty in gas metallicity in these relations is around $0.1-0.2$ dex. 
{The MZR$_{\rm gas}$ of a local compact galaxy population named Lyman-break analogues (LBAs), which resemble the relation at $z\sim1.5$ \citep{lian2015}, is also included for comparison.} 

It can be seen that the MZR$_{\rm gas}$ at high redshift is generally steeper than the local one. Most of the evolution of the {zero-point} of the MZR$_{\rm gas}$ seems to have happened at early times between $z\sim 3.5$ and $z\sim 1.5$ with a significant increase in gas metallicity roughly independent of galaxy stellar mass leaving the slope of the relation unchanged. Subsequently since $z\sim 1.5$ the gas metallicity of the most massive star forming galaxies has not been evolving any further, while most of the evolution is now happening in the lowest mass galaxies. The result is a considerable flattening of the relation between redshift $z\sim 1.5$ and $z=0$. This mass-dependent gas metallicity evolution has been already noted by \citet{lian2015} and \citet{kashino2017} but the physical explanation remains yet to be identified.  
The invariant gas metallicity in massive galaxies from $z\sim1.5$ to $z\sim0$ suggests that these galaxies are already efficiently enriched and their gas metallicity nearly saturates since $z\sim1.5$, while low-mass galaxies keep evolving in line with the widely accepted picture of downsizing in galaxy evolution 

\subsection{Comparison with stellar metallicities}
In Paper~I we derived stellar metallicity from SDSS galaxy spectra using the full spectral fitting code FIREFLY (\citealt{wilkinson2015,wilkinson2017}) and the stellar population models of Maraston and Str\"omb\"ack \citep{maraston2011} (with a Kroupa IMF). {The measurement of stellar metallicity is challenging owing to the well-known degeneracy between stellar age, stellar metallicity, and dust extinction \citep{worthey1994}. Therefore we provide a direct comparison of our results with metallicity measurements from the literature in Paper~I. We show that the MZR$_{\rm stars}$ used here is in good agreement with the relation in \citet{peng2015} based on measurements by \citet{gallazzi2005} and other determinations in the literature \citep{gallazzi2005,panter2008,thomas2010,johansson2012}. Also, the finding that the MZR$_{\rm stars}$ is steeper than the MZR$_{\rm gas}$ (see below) is well in line with the study by \citet{gonzalez2014} where the authors show that average stellar metallicities are generally lower than gas metallicties while the metallicities of the younger populations are higher and match the gas metallicity.}

Through the direct comparison between the gas and the stellar metallicities we then constrained the chemical evolution of our galaxy sample. 
We found that the MZR$_{\rm star}$ is much steeper than the MZR$_{\rm gas}$, implying a steeper MZR$_{\rm gas}$ at high redshift. 
This is qualitatively consistent with the finding of a steeper MZR$_{\rm gas}$ at high redshift as shown in Figure~1. { 
We note that \citet{ly2016} conclude not to find strong evidence for an evolution of the slope of MZR$_{\rm gas}$ since $z\sim 1$, which is as odds with other literature data shown in Figure~1. However, \citet{ly2016} focus on very low-mass galaxies with masses below $10^9\ M_{\odot}$, hence their data does not probe the mass-range within which evolution of the slope appears to occur.} 	  
{To understand the underlying physical processes that drive the evolution of the mass-metallicity relation, a detailed chemical evolution model is required that takes these processes into account.}

\section{The chemical evolution model}
To investigate the physical processes that drive the redshift evolution of the MZR$_{\rm gas}$, a full chemical evolution model is needed. In Paper I, we develop a numerical chemical evolution model and successfully design it to {\it simultaneously} reproduce both the MZR$_{\rm gas}$ and the MZR$_{\rm star}$ of local star-forming galaxies.

\subsection{Model parameters}
The model accounts for three basic physical processes that regulate the chemical evolution of galaxies, namely gas inflow, gas outflow and star formation. We assume a single gas inflow phase whose rate declines exponentially with time. 
The outflow is characterized by the mass fraction of stellar ejecta that are expelled from the galaxy which is proportional to the mass loading factor defined in the literature {as the ratio between the mass of the outgoing baryons and the star formation rate \citep{veilleux2005}}. 
The empirically-derived Kennicutt-Schmidt (KS, \citealt{kennicutt1998}) law is used to set the SFR at a given gas mass (surface density). 
In total, there are seven basic parameters in the model:
\begin{itemize}
\item The coefficient and exponent of the adopted KS law, $A_{\rm ks}$ (normalized to the original value and $n_{\rm ks}$,
\item The initial inflow rate, $A_{\rm inf}$ and its declining time scale $\tau_{\rm inf}$,
\item The outflow fraction $f_{\rm out}$,
\item The IMF slope at the low- and high-mass ends, $\alpha1$ and $\alpha2$,
\end{itemize}  
For more details about model calculations we refer  to \textsection3 in Paper I. 

\begin{figure*}
	\centering
	\includegraphics[width=2.1\columnwidth,viewport=10 10 1000 420,clip]{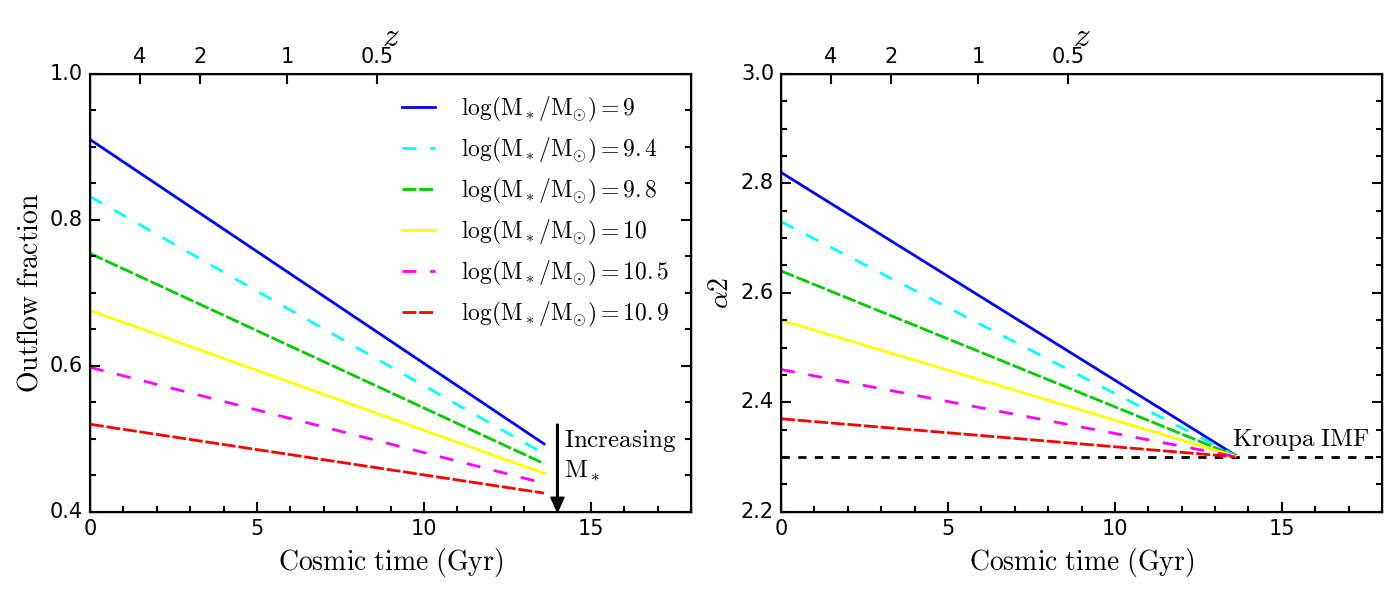}
	\caption{Time evolution of the metal outflow fraction (left-hand panel) and IMF slope at the high-mass end (right-hand panel) as a function of galaxy stellar mass adopted in the variable metal outflow and variable IMF models which will be shown in Figure 3.  
	}
	\label{figure2}
\end{figure*}
\subsection{Gas vs stellar metallicity}
In Paper I we extensively explore the parameter space to find the parameter combination that allows us to reproduce the observed gas and stellar metallicities of local star forming galaxies simultaneously. 
Since the gas metallicity is generally higher than the stellar metallicity, especially in low mass galaxies, 
the parameters that regulate gas and stellar metallicities need to be decoupled, and the metal production at early times needs to be suppressed. 
Since metal enrichment in the ISM is mainly driven by short-lived Type-II supernovae, the gas metallicity is regulated by very recent evolutionary processes. 
Stellar metallicity, instead, reflects the whole process of galaxy evolution. Therefore, an effective way of separating the drivers of gas and stellar metallicity is to impose a time dependence on the key processes affecting the metal enrichment.  

\subsection{Time-dependent metal outflow/IMF slope}
In Paper I we find that only two scenarios with either a time-dependent metal outflow fraction or a time-dependent IMF slope are successful in reproducing the observations. 
{We adopt a linear function to characterize the time evolution of the metal outflow fraction and the IMF slope. Figure~2 illustrates the time dependence of the metal outflow fraction (left-hand panel) and the IMF slope (at a star mass $M_*>0.5{\rm M_{\odot}}$, right-hand panel) as a function of galaxy stellar mass, according to the finding of Paper I. 
In the variable outflow scenario the metal outflow fraction decreases with time with higher initial metal outflow fractions and a stronger evolution in the least massive galaxies. Given the gas metallicity determined by the empirical N2 method, for a relatively massive, Milky Way-like galaxy ($M_*\sim6*10^{10}{\rm M_{\odot}}$), the metal outflow fraction is $\sim65\%$ and barely changes over a Hubble Time. In contrast, for a low mass galaxy like the Small Magellanic Cloud ($M_*\sim7*10^{9}{\rm M_{\odot}}$), the metal outflow fraction is high at early times (more than 80\%) and gradually decreases to $\sim70\%$ at the current epoch. The evolution of the IMF slope as a function of time and galaxy stellar mass follows a similar pattern (see right-hand panel of Figure~2).

In a nutshell, the former scenario manipulates {\em metal retention} in low-mass galaxies implying that {\em most metals are lost} from the galaxy at early times, while the latter manipulates {\em metal production} in low-mass galaxies implying that {\em less metals are produced} at early times.

It should be noted that the exact value of the metal outflow fraction depends on the adopted gas metallicity calibration. For example, with a gas metallicity determined through the R23 method \citep{pilyugin2005}, which typically leads to higher values than the empirical N2 method, the required metal outflow fraction is smaller. We refer the reader to Paper I for a detailed discussion of the dependence of the model parameters on the gas metallicity calibration.}

\subsection{The mass-metallicity relationships}
Based on these two scenarios we succeeded at reproducing simultaneously the MZR$_{\rm gas}$, the MZR$_{\rm star}$, and the mass-SFR relation of local star-forming galaxies. In a accompanying study  (Lian et al. 2018) we extended this model to also include spatial information and modelled the metallicity gradients of local star-forming galaxies observed with SDSS/MaNGA 
(Goddard et al. 2017). We showed that time-dependent metal outflow fractions or IMF slopes leading to the suppression of early metal enrichment at large radii are required to match gas- and stellar metallicity gradients simultaneously. 

In this work we focus on the redshift evolution. {The star formation law is assumed not to evolve with redshift.}
The adopted values for the parameters used in the variable metal outflow and variable IMF model flavours shown in Figure 3 are listed in Table 1. 
For simplicity we adopt a physical size of 5 kpc for all models independently of redshift. The effect of potential galaxy size evolution on the chemical evolution can be compensated by adopting different gas accretion time scales in order to keep the same gas surface density at any given redshift.   

\begin{figure*}
	\centering
	\includegraphics[width=18cm]{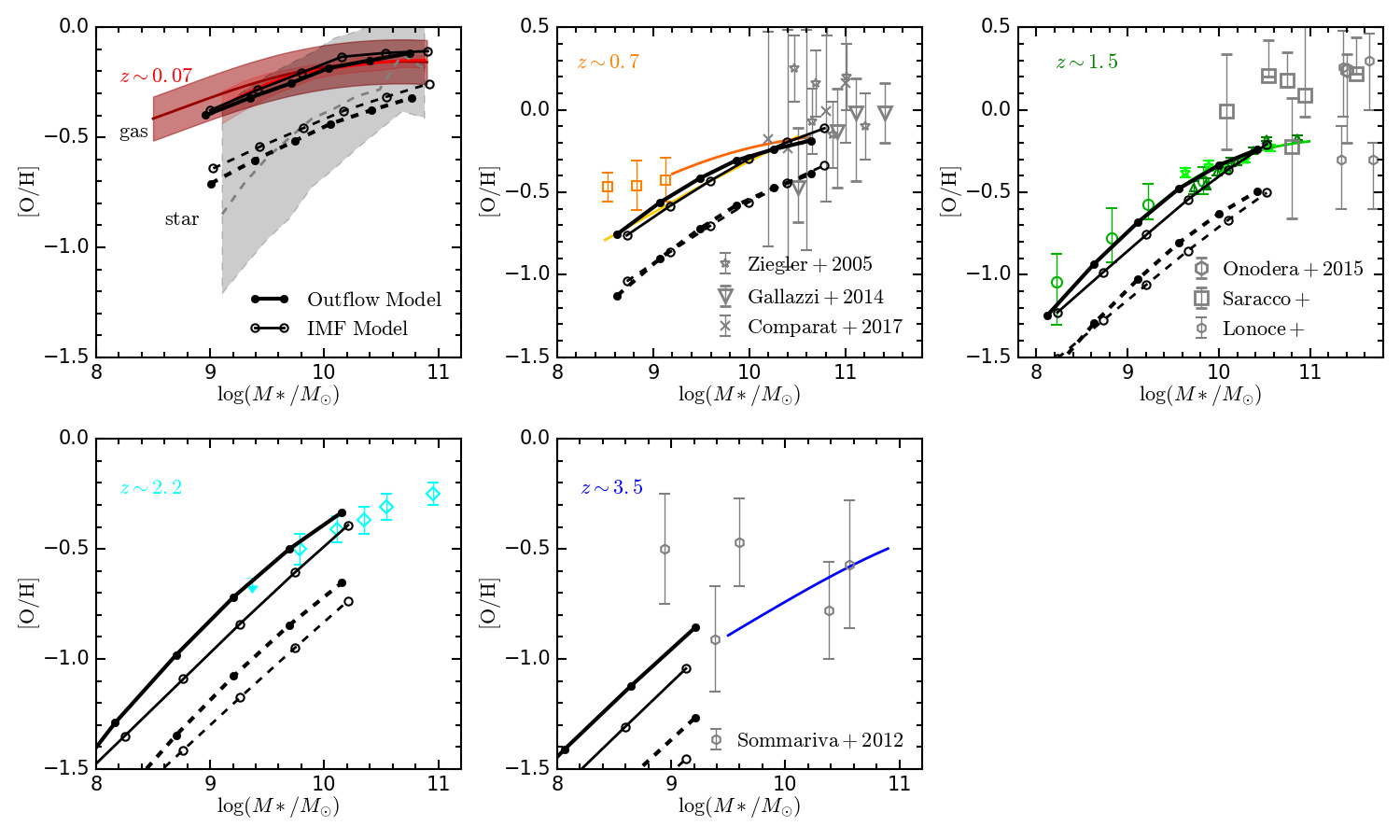}
	\caption{The mass-metallicity relation of star-forming galaxies for various redshift bins (see legend in the panels). Our models are shown as black solid (gas metallicity) and dashed (stellar metallicity) lines, with thick and thin lines referring to two model flavours (outflow and IMF model, see text for more detail). The coloured lines and symbols are observational data of gas metallicities following the colour scheme and symbol style of {Figure~1}. The observed mass-stellar metallicity relation of local galaxies from \citet{lian2018} is shown as gray dashed line with the gray area indicating the width of the distribution. {Other data for stellar metallicities of galaxies at high redshift are shown as gray symbols (see legend in the panels).}  
	}
	\label{figure3}
\end{figure*}

\begin{figure*}
	\centering
	\includegraphics[width=18cm]{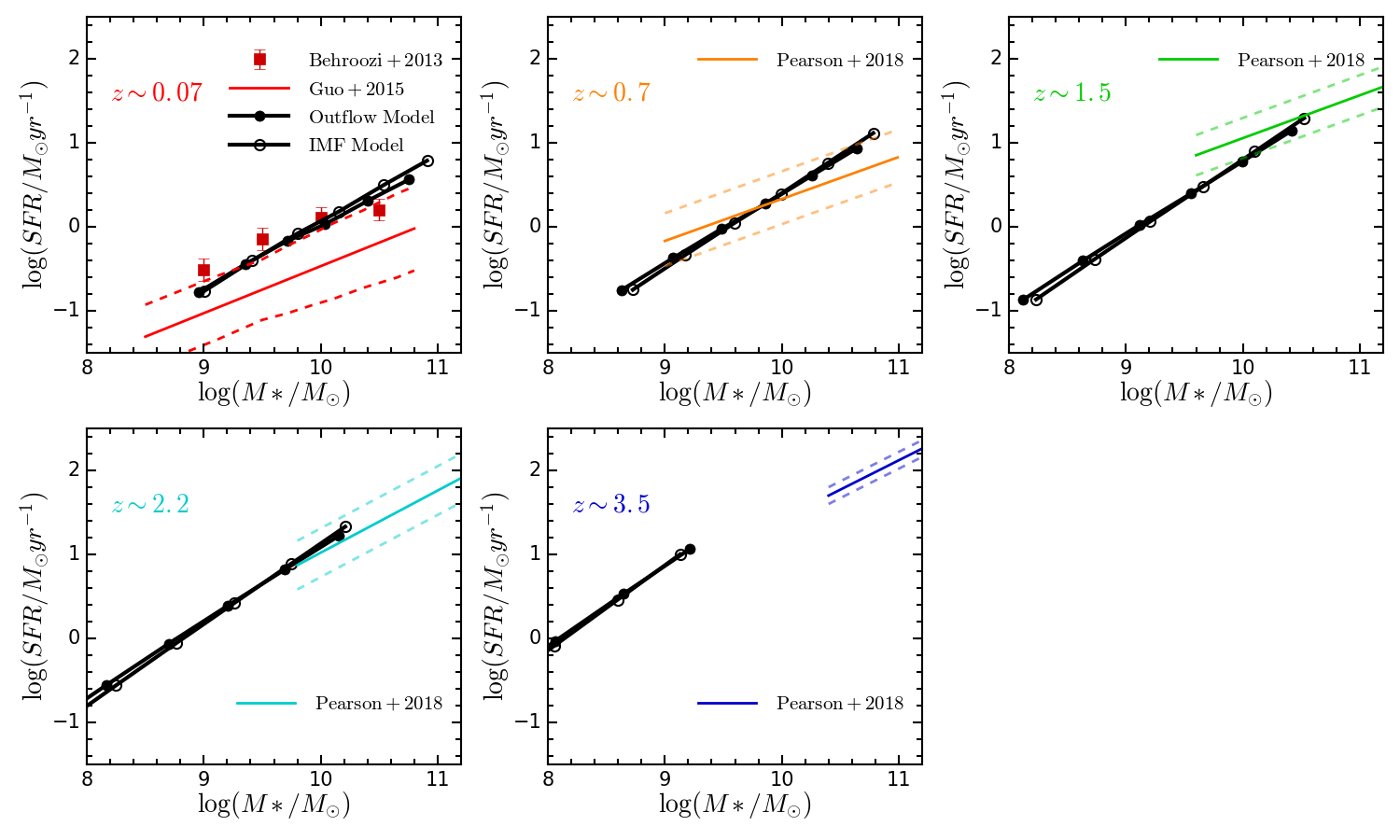}
	\caption{Comparison between the observations and prediction of chemical evolution models in the mass-SFR relation at different redshifts. Each panel indicates one redshift bin with the same colour scheme in Figure~3. Solid black lines connected by filled and empty circles indicate the prediction of models based on time-dependent metal outflow and time-dependent IMF scenario, respectively. The observed local mass-SFR relation are taken from \citet{behroozi2013,guo2015} while mass-SFR relations beyond local universe are taken from \citet{pearson2018}. Dashed lines and error bars show the scatter of the data distribution when applicable. }   
	\label{figure4}
\end{figure*}

\section{Results}
We compare the predictions of our models with the MZR$_{\rm gas}$ derived from data at various redshifts ($z\sim3.5$ to $z\sim0$) in Figure~3. The models of this work are shown as solid (gas metallicity) and dashed (stellar metallicity) black lines, with thick and thin lines referring to the two model flavours (outflow and IMF model). In order to construct the predicted MZR$_{\rm gas}$ and MZR$_{\rm star}$ (open and filled circles in Figure~3) we calculate models for six different final stellar masses. 

Coloured lines and symbols are observational literature results for gas metallicities following the colour scheme and symbol style of {Figure~1}. The observed mass-stellar metallicity relation of local galaxies from \citet{lian2018} is shown as {a} grey dashed line with the grey area indicating the {1$\sigma$ scatter} of the distribution. Grey triangles, hexagons, and stars in the upper middle panel are the observed stellar metallicities of early-type galaxies at $z\sim0.7$ \citep{ziegler2005,gallazzi2014,comparat2017,lonoce2019}. {Grey symbols in the upper right panel indicate the median stellar metallicity of samples of early-type galaxies at $z\sim1.6$ \citep{onodera2015,lonoce2015,saracco2017} while the hexagons in the bottom middle panel represent the observations of individual galaxies at $z>3$ from \citet{sommariva2012}.} 

\subsection{Gas metallicity evolution}
{It can be seen that the predictions of both the variable metal outflow and variable IMF model are nearly identical in this plane of gas metallicity vs stellar mass, and both match the redshift evolution of the MZR$_{\rm gas}$ very well.} 
As these same models also reproduce the local relations (Paper I), they 
naturally connect the gas metallicities observed in high redshift galaxies with the gas and stellar metallicities measured in galaxies locally. 

As already noted, the predictions of the two model flavours (time-dependent metal outflow vs time-dependent IMF slope) are degenerate in the parameter space shown in Figure~3. In other words, both mechanisms could be at work. For a detailed discussion of the pros and cons of these two models we refer to \citet{lian2018}. It is worth noting, though, that the required trend of the IMF as a function of galaxy stellar mass is opposite to the IMF variation argued for local massive quiescent galaxies in order to reproduce their near-IR spectral features (\citealt{conroy2012,cappellari2012,parikh2018}).

\subsection{Gas vs stellar metallicity evolution}
Our model also predicts the redshift evolution of MZR$_{\rm star}$ shown as dashed lines in Figure~3. It is interesting to note that the MZR$_{\rm gas}$ and MZR$_{\rm star}$ have very different slopes in the local universe, but similar slopes at high redshift with the MZR$_{\rm star}$ being significantly steeper than the MZR$_{\rm gas}$ today. In other words, the MZR$_{\rm gas}$ has experienced a stronger evolution in the shape than the MZR$_{\rm star}$ with the most significant {\em increase of gas metallicity in low-mass galaxies}. The predicted parallel slope of the MZR$_{\rm gas}$ and MZR$_{\rm star}$ at $z>1.5$ suggests that the physical drivers of the two relations are linked at early times and decouple at $z\sim1.5$, {i.e. past the epoch of maximum of the cosmic star formation rate}.

\subsection{Stellar metallicities at high redshifts}
There are not many observational constraints on the redshift evolution of the MZR$_{\rm star}$ due to the challenge of obtaining high quality spectra of distant galaxies with a reliable detection of the stellar continuum and absorption features. Existing samples in the literature often focus on passive, early-type galaxies or contain a mix of star-forming and passive galaxies. \citet{ziegler2005} obtain stellar metallicities for a sample of 13 early-type galaxies between redshift 0.2 and 0.4. They conclude that the stellar metallicities are well consistent with those of the local counterparts. \citet{gallazzi2014} push to somewhat higher redshifts and larger samples and derive the MZR$_{\rm star}$ for $\sim70$ star-forming and passive galaxies at $z\sim0.7$ with $M_*>10^{10}{M_{\odot}}$. They conclude that galaxies at $\sim0.7$ are systematically more metal poor in their stellar populations than their local counterparts at $\sim0.13$ dex. {\citet{comparat2017} derive stellar population properties for a large sample of intermediate redshift galaxies, including star forming emission-line galaxies and luminous red galaxies, from the Extended Baryon Oscillation Spectroscopic Survey, eBOSS \citep{dawson2016,blanton2017,abolfathi2018} using full spectral fitting with FIREFLY.} \citet{lonoce2015,lonoce2019} present individual measurements for a sample of four massive galaxies between redshifts $z\sim 1.4$ and $z\sim 2$, and \citep{onodera2015} provides a measurement of stellar metallicity for a stacked spectrum of a sample of galaxies around this same redshift. Metallicity measurements of further six early-type galaxies in this redshift bin are provided by Saracco et al (in prep) selected from their sample presented in \citet{saracco2017}.

{All these} studies show that massive galaxies were already metal-rich and dominated by old stellar populations at those early epochs. Finally, \citet{sommariva2012} derive stellar metallicities for a sample of star-forming galaxies at $z>3$ based on UV absorption features and found the stellar metallicities in these galaxies to be broadly consistent with the gas phase ones.  

The data is shown in Figure~3 as open grey symbols. Most stellar metallicities at high redshift are higher than predicted by our model (dashed lines in extrapolation). This ought to be expected since the observed high-$z$~stellar metallicities refer to quiescent, early-type galaxies, for which the stellar metallicity is generally higher than in star forming galaxies \citep{peng2015}. The only exception is the lowest-stellar mass data point by \citet{gallazzi2014} at $z\sim 0.7$, which is likely the only one dominated by star-forming galaxies. The Gallazzi et al. results imply significantly lower metallicities in the stars than in the gas also in galaxies at redshift $z\sim 0.7$ in good agreement with the prediction from our model. 

{Some of the measurements in \citet{sommariva2012} are higher than the prediction of our model at $z>3$ which may be due to the different stellar populations represented by the data and the model. \citet{sommariva2012} use $UV$ absorption features to derive the stellar metallicity, which gives more weight to the properties of the youngest, most metal-rich stars. The mass-weighted stellar metallicity predicted by our model, instead, represents the average value over all ages of the stellar population.} 

\subsection{The mass-SFR relationship}
{We further examine our model by comparing the predicted mass-SFR relationship (i.e. main sequence, \citealt{elbaz2007,noeske2007,renzini2015}) with observations at the various redshift intervals. The result is shown in Figure~4. Each panel indicates one redshift bin with the same colour scheme as in Figure~3. Also the symbols and line styles are the same as in Figure~3. The observed local mass-SFR relation is adopted from \citet{guo2015} {based on H$\alpha$ in combination with 22$\mu$m observations from WISE and} \citet{behroozi2013} {with SFRs adopted from \citet{salim2007} and \citet{robotham2011} based on UV observations with {\sl GALEX}}. The mass-SFR relations beyond the local universe are taken from a recent work by \citet{pearson2018} based on deep Herschel far-IR observations. Dashed lines and error bars indicate the $1\sigma$ scatter of the data distribution. {The local SFR measurements are broadly consistent within the 1$\sigma$ scatter. The difference may be due to the difference in sample selection of star-forming galaxies and usage of different SFR indicators which trace the SFR over different time scales. The SFR measured by mid-IR observations in the local universe should in principle be consistent with the SFR based on far-IR observations at high redshift given the good correlation between mid- and far-IR luminosity with the total IR luminosity of star forming galaxies (e.g. \citealt{rieke2009}).} 

It can be seen that the mass-SFR relations at different redshifts predicted by our chemical evolution model are in generally good agreement with the observations. The predicted slope is slightly flatter than the one observed at intermediate redshifts, but still consistent within the $1\sigma$ error. A further adjustment of the model parameters controlling the star formation efficiency and the inflow/outflow rates can mitigate the discrepancy, which, however, goes beyond the scope of this paper.
}

\begin{table*}
	\caption{Parameter ranges of the time dependent chemical enrichment models (for a variable outflow and a variable IMF) shown in Figure 4.}
	\label{table5}
	\centering
	\begin{tabular}{l c c c c c c c c c c}
		\hline\hline
		& $A_{\rm ks}$ & $A_{\rm ks,i}$ & $n_{\rm ks}$ & $A_{\rm inf}$ & $\tau_{\rm inf}$ & $f_{\rm out}$ & $f_{\rm out,i}$ & $\alpha1$ & $\alpha2_{\rm i}$ & $\alpha2$ \\
		& & & & ${\rm M_{\odot} yr}^{-1}$ & Gyr & & & & \\
		\hline
		Variable outflow & 1 & 0.63 & 1.5 & [0.20,19.95] & [40,6.6] & [0.50,0.43] & [0.92,0.50] & 1.3 & 2.3 & 2.3 \\
        Variable IMF & 1 & [1,0.50] & 1.5 & [0.20,25.12] & [40,6.6] & [0.40,0.35] & [0.40,0.34] & 1.3 & [2.85,2.3] & 2.3 \\		
		
		\hline
	\end{tabular}\\  
\end{table*} 

\section{Discussion} 
As shown in Figure~3 the MZR$_{\rm gas}$ has been flattening with time since $z\sim 1.5$. Massive star-forming galaxies reach their final gas metallicity already at that redshift, while the gas metallicity in low-mass galaxies kept increasing steadily until today. This chemical downsizing pattern \citep{maiolino2008} imprinted in the metal-enrichment histories of star-forming galaxies is similar, but not the same as the downsizing seen in the stellar population properties of passive galaxies \citep[e.g.,][]{cowie1996,heavens2004,thomas2005,thomas2010}. The latter is attributed to quenching and caused by the fact that more massive galaxies quench earlier. The chemical downsizing discussed here, instead, is detected in a population of galaxies forming stars throughout, and must therefore be attributed to the process of metal enrichment itself. 

\subsection{The origin of the mass-metallicity relation}
The redshift evolution of the MZR$_{\rm gas}$ requires two separate mechanisms to be at play. The metal enrichment 1) needs to be suppressed at early times in low-mass galaxies, and 2) must come to a halt in massive galaxies at an epoch around $z\sim 1.5$.
A time-dependent mechanism that regulates {metal enrichment} in galaxies is required. One possibility is to invoke a time-dependence in the star formation efficiency (SFE), predominantly in lower-mass galaxies, such that the SFE increases with cosmic time. 
Based on an analytical galaxy evolution model, \citet{lilly2013} investigate the cosmic evolution of the MZR$_{\rm gas}$ and show that a decline in SFE with time can drive the evolution of the MZR$_{\rm gas}$. 
\citet{sakstein2011} come to a similar conclusion based on a semi-analytical model showing that the same parameters that regulate the SFR are also responsible for the evolution of MZR$_{\rm gas}$. 
However, as discussed in Paper I, this time-dependent SFE scenario fails at reproducing simultaneously the MZR$_{\rm gas}$ and MZR$_{\rm star}$ in the local universe.  
In such a model the predicted MZR$_{\rm star}$ evolves in parallel to the MZR$_{\rm gas}$ leading to an MZR$_{\rm star}$ that is too shallow compared to observations, as shown in Figure~3 and discussed in Paper~I.

\subsection{Time-dependent metal outflow vs IMF slope}
With the present study we present a model that matches both, the local mass-metallicity relation of stars and gas, as well as the redshift evolution of the MZR$_{\rm gas}$. This model is based on a time-dependent metal outflow loading factor or a time-dependent IMF slope rather than a time-dependent star formation efficiency. In both these model flavours, the production of metals is substantially suppressed in low-mass galaxies at early times, which explains both the low stellar metallicity in today's low-mass galaxies and the low gas metallicity in high-redshift low-mass galaxies.


{It is plausible to assume that a shallower gravitational well at earlier times leads to a higher metal loading factor in galactic winds. This naturally leads to a dependence on cosmic time, galaxy mass, and galaxy radius as inferred in our model. We note that such mass dependence of the metal loading factor is also required by the chemical evolution model by \citet{peeples2011} to explain the observed correlation of gas metallicity and gas fraction with galaxy mass. The time-dependent IMF scenario, instead, could be caused by a metallicity dependence of the IMF if steeper IMFs are generated in lower metallicity environments. This is not implausible, but clear observational evidence for such a behaviour is currently missing \citep{bastian2010,bate2014} and opposite trends seem to exist in simulations \citep{kroupa2013,chabrier2014}.}	

To summarise, the redshift evolution of the MZR$_{\rm gas}$ is driven by a combination of mass-dependent star formation efficiency and a time-dependent suppression of metal enrichment in low-mass galaxies through either efficient metal outflow or a steeper IMF slope at early times.

\subsection{Gas metallicities in massive galaxies}
It is interesting to note that the gas metallicity in massive star-forming galaxies stops increasing with cosmic time at around $z\sim 1.5$ despite ongoing star formation activity. This saturation in the metal enrichment is predicted naturally by any chemical evolution model.
{The reason is that the metal enrichment in the gas is the result of competing effects between star formation-driven metal production and dilution caused by accretion of pristine gas. In the low metallicity regime, the dilution effect is inefficient compared to the star formation-driven metal production.  
Once the gas metallicity becomes high enough, the dilution effect due to gas accretion becomes efficient.}
{Therefore, saturation} occurs when the gas metallicity reaches a high enough value {for the dilution effect to balance the metal production}. At this point the metallicity settles at the effective yield of metal production \citep{edmunds1990,thomas1999}.
{It is interesting to note that the saturation time for massive star forming galaxies ($M_*>10^{10.5}{\rm M_{\odot}}$) is very close to the peak of the cosmic star formation history ($z\sim2$, \citealt{madau2014}).}

This saturation feature is also evident in Figure~12 in Paper I where we explore all possible cosmic evolution of gas and stellar metallicity under various model parameter configuration,   
and is a natural explanation for the change of evolution mode of the MZR$_{\rm gas}$ at $z\sim1.5$ \citep{lian2015}. At redshift $z>1.5$, the gas metallicity is still far from the threshold value and metal enrichment is still efficient also in massive galaxies. As a consequence only the zero-point, but not the slope of the MZR$_{\rm gas}$ evolves from $z\sim3.5$ to $z\sim1.5$. Since more massive galaxies generally have higher gas mass surface density and therefore higher efficiency in converting gas into stars and metal production, they reach the saturated metallicity earlier than less massive galaxies. At $z\sim1.5$, the gas metallicity in massive galaxies ($M_*>10^{10.5}\ {\rm M_{\odot}}$) has first reached the threshold value and saturates. As a result, the high mass end of MZR$_{\rm gas}$ remains unchanged after $z\sim1.5$ and the MZR$_{\rm gas}$ becomes flatter with time. {It should be noted that the stellar mass of a galaxy is increasing along with the metal enrichment process. Therefore, the evolution in the mass-gas metallicity relation indicates that the metal enrichment of galaxies progresses at a faster pace than their mass assembly until they reach the threshold metallicity of saturation.}
{Since the saturation value of the gas metallicity depends on the IMF, the invariant gas metallicity of the massive galaxies ($M_*>10^{10.5}\ {\rm M_{\odot}}$) from $z\sim1.5$ to $z\sim0$ serves as a strong evidence for an invariant IMF in those galaxies in this redshift range.} 


\subsection{Cosmological models}
{Based on semi-analytical models and cosmological hydrodynamical simulations, \citet{guo2016} found the cosmic evolution of the MZR$_{\rm gas}$ to be inconsistent between these two theoretical approaches. A flattening of the MZR$_{\rm gas}$ and with increasing gas metallicity over cosmic time is predicted by the hydrodynamical simulations, which is more consistent with the observations and the chemical evolution model presented here. The lack of evolution predicted by the semi-analytical models may be due to a deficient treatment of gas flows \citep{guo2016}. 

A similar trend of a flattening of the MZR$_{\rm gas}$ with cosmic time is also found by \citet{taylor2016} based on their cosmological hydrodynamical simulations. Interestingly, their simulation further shows that the MZR$_{\rm star}$ mainly evolves in zero point and not slope in contrast to the MZR$_{\rm gas}$. This prediction by the model is in good qualitative agreement with the prediction of our numerical chemical evolution model. Although the state-of-art cosmological hydrodynamical simulations seem to predict a cosmic evolution trend of MZR$_{\rm gas}$ broadly consistent with observations, it should be pointed out that they still face difficulties to match the exact shape of the observed MZR$_{\rm gas}$ at different redshifts (see Figure~12 in \citealt{guo2016} and Figure~8 in \citealt{taylor2016}).}



\section{Conclusions}
We study the origin and cosmic evolution of the mass-metallicity relation (MZR) in star-forming galaxies based on a full, numerical chemical evolution model which was designed to match the local MZRs for both gas and stars simultaneously (Lian et al. 2018, Paper I). 

Our model accounts for three basic physical processes that regulate the chemical evolution of galaxies, namely gas inflow, gas outflow, and star formation.
The main characteristic of this model is that a time-dependent metal enrichment process is invoked assuming either a time-dependent metal outflow with {higher} metal loading factors in galactic winds at early times, or a time-dependent IMF with steeper IMF slopes at early times.
{The former regulates the {\em metal retention} in galaxies so that more metals are lost at early times, while the latter regulates the {\em metal production} so that less metals are produced at early times.} 
The aim of the present paper is to test the predictions by these models for the redshift evolution of the MZR$_{\rm gas}$ and compare them with observational data at various redshifts up to $z\sim3.5$. 

To this end we analyse our own sample drawn from SDSS and discussed in Paper~I and adopt further data sets from the literature covering the redshift range $0\leq z\leq 3.5$. We homogenise the sample by correcting stellar masses to the stellar mass based on the Kroupa IMF \citep{kroupa2001}, and gas metallicities to the empirical N2 method \citep{pettini2004}. The data suggest a two-phase evolution with a transition point around $z\sim 1.5$. Before that epoch the MZR$_{\rm gas}$ has been evolving parallel with no evolution in the slope. After $z\sim1.5$ the MZR$_{\rm gas}$ started flattening until today. 
Hence the gas metallicity of the most massive star forming galaxies has not been evolving since that epoch. This is an epoch just past the peak of the cosmic star formation rate \citep{madau2014}. Instead, significant evolution has been occurring in the lowest mass galaxies until today. 

This is in good agreement with the steep MZR$_{\rm star}$ of local star-forming galaxies, which indeed implies a steep MZR$_{\rm gas}$ at early times. 
%
This delayed-enrichment model matches the redshift evolution of the MZR$_{\rm gas}$ very well.
Hence we have a working model which naturally connects the gas metallicities observed in high redshift galaxies with the gas and stellar metallicities measured in galaxies locally. {Our model also reproduces the evolution of the main sequence, hence the correlation between galaxy mass and star formation rate.}

We further discuss the main drivers of the mass-metallicity relation and its evolution with redshift. It turns out that the redshift evolution of the MZR$_{\rm gas}$ requires two separate mechanisms to be at play. The metal production needs to be suppressed at early times in low-mass galaxies, and must come to a halt in massive galaxies at an epoch around $z\sim 1.5$. Our model invoking a time-dependent mechanism that regulates star formation in galaxies through either a time-dependent metal outflow or a time-dependent IMF slope matches this behaviour very well.  

Our model also predicts the redshift evolution of the MZR$_{\rm star}$. There are not many observational constraints on the redshift evolution of the MZR$_{\rm star}$ due to the challenges in the observations. We compare the prediction of our model with data from the literature. The latter mostly contains data of massive, quenched early-type galaxies, while here we model the evolution of star-forming galaxies. As a consequence most of the stellar metallicities at high redshifts are higher in the data than predicted by our model as expected. 

Data of stellar metallicities of lower-mass ($<10^{11}\;M_\odot$), star-forming galaxies at high redshift is needed for a meaningful test of our model.

\section*{Acknowledgements}
The Science, Technology and Facilities Council is acknowledged for support through the Consolidated Grant ‘Cosmology and Astrophysics at Portsmouth’, ST/N000668/1. This work is also supported by the National Natural Science Foundation of China (no. 11673004). Numerical computations were done on the Sciama High Performance Compute (HPC) cluster which is supported by the ICG, SEPnet and the University of Portsmouth.

\end{document}